# 3D FRONSAC with PSF reconstruction

Yanitza Rodriguez[1], Nahla M. H. Elsaid[1], Boris Keil[2], and Gigi Galiana[1]

[1]Department of Radiology and Biomedical Imaging. Yale School of Medicine, New Haven, CT, United States

[2]Institute of Medical Physics and Radiation Protection, Department of Life Science Engineering, TH Mittelhessen University of Applied Sciences, Giessen, Germany

1 | INTRODUCTION

Many strategies to improve parallel imaging can be categorized as either (1) more advantageous paths through k-space [1], [2], (2) better reconstruction [3], [4], [5], [6], or (3) improved receiver hardware [7], [8], [9], [10]. An emerging strategy is to apply nonlinear gradients [12], [20], especially with highly dynamic waveforms [14], [15], [16]. In particular, the FRONSAC trajectory, which applies a modest amplitude but rapidly oscillating waveform on several nonlinear gradient channels, has been shown to provide very good image acceleration while maintaining the features of the underlying linear gradient trajectory.

One particularly advantageous trajectory is Cartesian-FRONSAC, where FRONSAC nonlinear encoding is applied to a traditionally undersampled Cartesian trajectory. Both simulations and experiments have shown that Cartesian-FRONSAC, unlike many non-Cartesian trajectories, maintains many of the attractive features that have kept Cartesian sampling as the workhorse of clinical imaging [16]. For example, the method is similarly insensitive to off-resonance spins, gradient timing errors, or contrast that evolves during an acquisition, such as that of a fast spin-echo acquisition. However, Cartesian-FRONSAC showed very good image quality with high undersampling factors (R=8 with an 8channel coil), which is generally not observed with Cartesian sampling. Work in our group and others has shown that FRONSAC performance was similar whether the nonlinear gradients were based on mathematical functions or more hardware-driven shapes, like those created by matrix coil arrays [12], [16], [17].

There are several ways to describe how nonlinear gradients improve the conditioning of undersampled image reconstruction. Just as RF coil sensitivities can be described as changing the pointwise sampling of k-space measurements of distributions of k-space, nonlinear gradient moment similarly smears the sampling function in k-space as a function of what shapes and moments are applied [18], [19]. This smearing of k-space into regions not sampled by the nominal gradient trajectory is the heart of what allows parallel imaging to infer k-space points that are nominally skipped in the undersampled trajectory. However, each RF coil modifies the k-space sampling function according to its sensitivity profile, which is usually fixed throughout an acquisition. In contrast, nonlinear gradient moment allows these sampling functions to be modulated rapidly during the readout. This additional degree of freedom, combined with oversampling in the readout direction, can be used to acquire staggered sampling functions at different orientations and with varying extents in k-space, providing additional information to fill gaps in the nominal k-space trajectory. This sampling strategy is the logic behind FRONSAC encoding.

Alternately, for Cartesian-FRONSAC, one can describe the rapidly oscillating gradient as modifying the point spread function (PSF) of each voxel along the readout direction, similar to the interpretation conventionally presented for wave encoding [1]. One crucial difference with nonlinear gradients, which are typically functions of multiple coordinates, is that the oscillating phase is different at every pixel, so there is a different PSF for every pixel in the image, not just every row. For each voxel, the rapidly oscillating nonlinear gradient creates a distinctive pattern of ghosts along the oversampled read direction. Because aliased voxels have unique ghosting patterns, the ability to unfold these voxels is improved over coil modulation alone. (Figure 1)

This PSF perspective of FRONSAC encoding can have significant advantages for reconstruction, particularly for high resolution or 3D acquisitions. In previous FRONSAC reconstructions, data and encoding were handled in the time domain. Encoding across the entire volume is coupled to the entire dataset, leading to substantial inversion problems. (Figure 2) However, as shown previously [1], a phase modulation that occurs during every readout of a frequency-phase encoded acquisition manifests as a PSF modulation in the readout direction of the Fourier transformed data. Thus, one line in the Fourier transformed data, aggregated across coils, relates only to R aliased lines in the reconstructed image volume, as illustrated in Figure 2. This is equivalent to block diagonalizing the encoding matrix into a series of far smaller problems, leading to dramatic efficiency gains.

Previous work has demonstrated strong advantages in undersampled image quality for various 2D Cartesian FRONSAC acquisitions using the full volume encoding matrix. This work presents the first examples of FRONSAC incorporated into a 3D Cartesian acquisition, with human images of both gradient echo and MPRAGE scans. Results are shown for both 8channel and 32channel receivers. The results show that FRONSAC encoding does not significantly modify the contrast obtained in either sequence, but it significantly improves undersampled reconstructions.

2 | THEORY

The PSF-based reconstruction of Cartesian-FRONSAC data generalizes the reconstruction described for wave-CAIPI [1]. Considering an arbitrary nonlinear encoding where L gradient channels are played simultaneously. The spatial variation of the $l$th gradient is $\Psi_l(x, y, z)$ and its temporal waveform is $A_l(t)$. Thus, the accumulated gradient moment is $k_l(t) = \int_0^t A_l(\tau)d\tau$ and the net modulation imparted by all L nonlinear gradient channels is $P_{nlg}$:

$$P_{nlg}(x,y,z,t) = \sum_{l=1}^{L} k_l(t)\Psi_l(x,y,z) \quad (1)$$

The Cartesian-FRONSAC signal can be described as the sum of a rectilinear encoding described by $k_x$, $k_y$, and $k_z$ and the arbitrary nonlinear encoding which is a function of time:

$$S(t, k_y, k_z) = \int m(x,y,z)exp(i2\pi[k_x(t)x + k_y y + k_z z + P_{nlg}(x,y,z,t)])dxdydz \quad (2)$$

Since $P_{nlg}$ is independent of $k_y$ and $k_z$, inverse Fourier Transform(iFT) with respect to $k_y$ and

$k_z$ yields:

$$S(t,y,z) = \int m(x,y,z)exp(i2\pi[k_x(t)x + P_{nlg}(x,y,z,t)])dx \quad (3)$$

For a Cartesian trajectory, $k_x = G_x t$, so the signal can also be inverse Fourier transformed with respect to t yielding:

$$S(\omega,y,z) = \int m(x,y,z)[iFT(exp(i2\pi P_{nlg}(x,y,z,t))) \otimes (\gamma(\omega - G_x x))]dx \quad (4)$$

Along the readout dimension, m(x,y,z) becomes a weighting on a PSF centered at the typical frequency expected from the linear gradient encoding. The shape of the PSF reflects the frequency content of the phase modulation brought about by the nonlinear gradients:

$$PSF_{(x,y,z)}(\omega) = iFT(exp(i2\pi P_{nlg}(x,y,z,t))) \quad (5)$$

For FRONSAC gradients with a single frequency, this is a series of harmonics related to the waveform frequency, where the amplitude and phase of each harmonic peak varies by pixel. While this PSF is generally centered in the bandwidth of the linear encoding (assuming no linear components in $P_{nlg}$), it can extend across a much wider spectrum, which is captured by oversampling in the readout direction. This additional encoding in the readout dimension improves the separability of signals folded along y and z, which is the heart of the improved parallel imaging from FRONSAC. (Figure 1)

For a multicoil acquisition, the PSF encoding matrix for a set of folded y, z locations, as acquired by multiple coils is:

$$E_{PSF}m = d$$

Where $E_{PSF}$ is a matrix of dimensions $(N_{timepoints}N_{coil}) \times (N_x R_y R_z)$, where $N_{timepoints}$ is the size of the oversampled readout, $N_{coil}$ is the number of receive coils, $N_x$ is the reconstruction size along readout dimension and $R_y$, $R_z$ are undersampling factors in each phase encode dimension. m is the true magnetization density, and d is a vector that stacks $S(\omega, y, z)$ acquired by each coil. The columns comprise all the locations that must be solved simultaneously, i.e. the $R_y R_z$ aliased lines of the image volume. Each column of $E_{PSF}$ is $N_{coil}$ copies of the PSF for that location, with each copy weighted by the location's coil sensitivity. Compared to a full volume reconstruction, the number of equations to be inverted is equal to the number of phase encode lines, but each encoding matrix is smaller by a factor equal to the square of the total phase encode lines, gaining significant reconstruction efficiency.

3 | METHODS

Data were acquired on healthy volunteers using either 8channel or 32channel RF coil arrays nested in a 1ch Tx/Rx coil. Experiments were performed at Yale University using a 3T MRI scanner (MAGNETOM Trio Tim, Siemens Healthcare, Erlangen, Germany). All data were acquired with linear gradients corresponding to $128^3$ Nyquist sampling for a bandwidth 130Hz/pix, but data was acquired with 8-fold oversampling in the readout to fully capture the phase modulations induced by FRONSAC encoding. Parameters for 3D GRE human experiments were as follows: TR=25 ms; TE=10 ms; flip angle=10°; FOV = $(250mm)^3$; bandwidth 130Hz/pix. Parameters for 3D MPRAGE human experiments were: TR=25 ms;

TE=10 ms; TI=1600ms; TD=500ms; flip angle=10°; acquisition matrix and linear gradients as in GRE experiments. In addition, read direction was chosen to be HF, to allow for 2D undersampling even with the 8channel azimuthal coil.

FRONSAC experiments added oscillating nonlinear gradients to the readout gradient, with no other changes to the acquisition. The gradient was produced using an insert head gradient (Tesla Engineering Ltd, Storrington, UK) rated at 321A with an inner diameter of 380mm which generates 3 spherical harmonic gradient fields: $x^3-3xy^2$, $3yx^2-y^3$ and $x^2+y^2$ (commonly known as C3, S3, and Z2). The gradient coil is capable of achieving maximum C3, S3, and Z2 fields of 3255mT/$m^3$, 3155mT/$m^3$ and 475mT/$m^2$, respectively. All studies used the same FRONSAC waveform of 64 cycles per readout (8.32kHz) on gradient channels C3, S3, and Z2, with amplitudes 447mT/$m^3$, 468mT/$m^3$, and 64mT/$m^2$, in triangular analogs to sine, cosine, and sine, respectively.

Nonlinear gradient trajectories were measured with a phase mapping sequence which is analogous to chemical shift imaging, as previously described [20]. 3D mapping data was acquired at half the image resolution with a more extended FOV and matrix to accommodate different patient positions (375x250x250$mm^3$ and 192x128x128 matrix). Phase images were spatially unwrapped and high SNR voxels of each time point were fit to a 6th order polynomial. These coefficients were then used to generate the nonlinear phases used in the reconstruction.

B1 maps for each subject were generated from the multichannel GRE images along with an identical single-channel acquisition. Multichannel acquisitions were normalized for magnetization density using single-channel acquisition, and the resulting B1 sensitivities were masked and fit to a smoothed polynomial. All images were undersampled retrospectively and reconstructed in MATLAB (MathWorks Inc, Natick, Massachusetts, USA) via inversion of the PSF matrix [21] described in the Theory section. 3D experiments were reconstructed at the rectangular dimensions of the field maps and trimmed to a final matrix of size $128^3$.

## 4 | RESULTS

Figure 3 shows three views of fully sampled 3D volumes acquired with and without FRONSAC encoding. Because FRONSAC encoding provides just a small perturbation to the Cartesian encoding trajectory, it delivers similarly timed traversal of k-space, which results in similar image quality and contrast. This figure demonstrates the consistent contrast observed between standard Cartesian and Cartesian-FRONSAC encoding for both GRE and MPRAGE acquisitions and contrast.

Figure 4 shows undersampled reconstructions of 32channel image acquisitions. In the 2nd and 4th columns, difference images are taken relative to the fully sampled Cartesian reconstruction, giving an inherent advantage to the non-FRONSAC images, which are undersampled versions of the same dataset. Despite this, visual inspection, structural similarity index, and NRMSE are consistently better in the undersampled FRONSAC images. Notably, even in these unregularized reconstructions, the undersampling artifacts in the 2x4 and 4x2 reconstructions are relatively modest ghost artifacts. These artifacts may be amenable to further reduction via more sophisticated reconstruction approaches. Figure 5 also shows reconstructions at various undersampling factors but for acquisition with an 8channel azimuthal coil. In this case also, visual inspection, SSIM and NRMSE are consistently better in the Cartesian-FRONSAC images. Interestingly, even for this low channel count, very highly accelerated reconstructions (4x2 and 2x4) show similar undersampling artifacts to the 32channel case.

## 5 | DISCUSSION

These results are the first to demonstrate experimental 3D FRONSAC imaging, and they show that the improvements in image quality are similar to those seen in 2D imaging. This method does require additional hardware, which is the main obstacle to its implementation at other sites. However, recent work developing gradient arrays for shimming is opening opportunities for the unique spatial encoding that can be delivered with this kind of hardware [15], [17], [22], [23], [24]. As FRONSAC gradients demand high slew rates but modest amplitudes, such devices would be well suited for implementing these kinds of encodings.

The use of FRONSAC encoding also requires careful characterization of the nonlinear gradient waveform, which is performed here with a method that creates a phase map for each timepoint of the acquisition. While the presented results were taken at uniform scan parameters, it is expected that, as shown in 2D, the same FRONSAC waveform can be used to improve images with different contrast, resolution, FOV and orientation. Therefore, a single field mapping can be used to improve a large range of acquisitions.

This work is also the first to describe FRONSAC encoding as a PSF modulation, which has several implications. As exploited here, the PSF formulation allows for far more efficient reconstruction strategies, making 3D reconstructions practical. PSF-based reconstruction as described for wave-CAIPI can generalized to a wide range of frequency+phase based encoding schemes with a dynamic modulation that is repeated for each readout [26]. The gradients used for any of those com ponents can, in principle, be spatially nonlinear, as is the case for Cartesian-FRONSAC. However, other FRONSAC methods (e.g., spiral-FRONSAC or rosette-FRONSAC) and projection based encodings (e.g., O-Space) are not directly amenable to this interpretation [20], [27]. Whether a PSF reconstruction approach can be extended to such techniques, perhaps by modifying the dynamic waveform to achieve some desired regularity in k-space, is an open question. PSF description of FRONSAC encoding also allows a more local interpretation of how a given waveform translates into enhanced parallel imaging, as demonstrated in Figure 1. This could be used to create optimized waveforms even for hardware with many arbitrarily shaped channels, such as gradient arrays.

## 6 | CONCLUSION

This work demonstrates that Cartesian-FRONSAC, which was previously demonstrated in 2D acquisitions, provides similar performance in 3D acquisitions. Because FRONSAC adds a relatively small perturbation to the spatial encoding, it does not significantly affect contrast, even when magnetization is evolving during acquisition (e.g. MPRAGE). However, Cartesian-FRONSAC encoding yields highly undersampled reconstructions that are more promising, with consistently better SSIM and NRMSE. Furthermore, it is demonstrated that FRONSAC, like wave-CAIPI, can be described as a PSF modulation which leads to more efficient reconstruction approaches.

## 7 | ACKNOWLEDGMENTS

The authors are grateful for support from Andrew Dewdney of Siemens and Terry Nixon of Yale with implementing the nonlinear gradient hardware. This work was funded by NIBIB R01 EB022030.

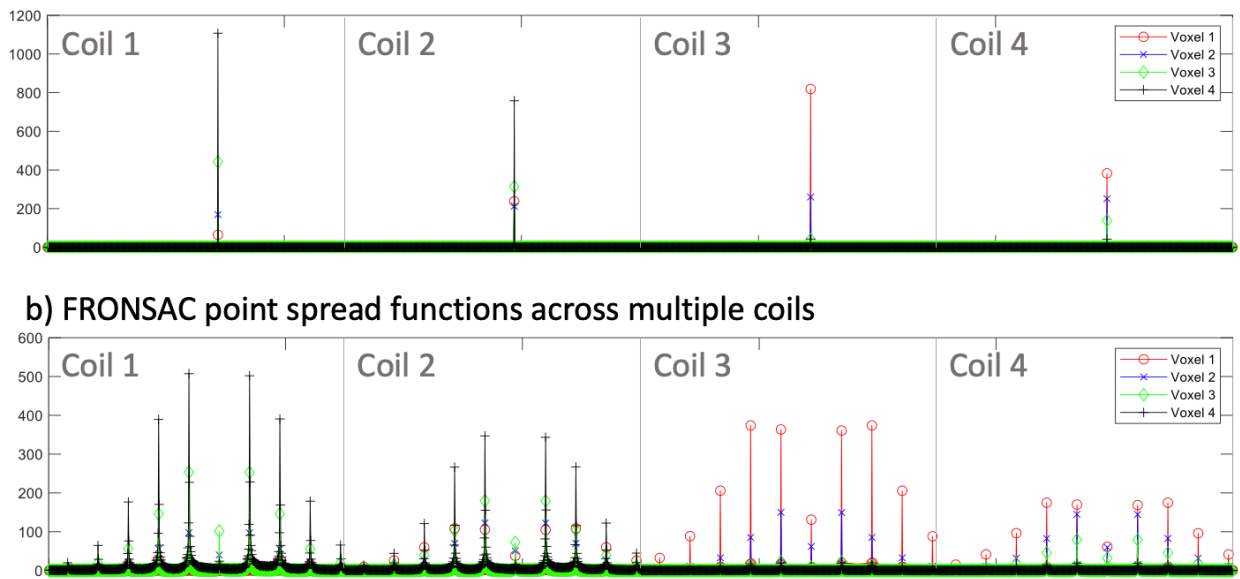

Figure 1: Point spread functions (magnitude only, cropped along the frequency axis to improve visualization) for four aliasing voxels when acquired by 4 different coils are shown for (a) Cartesian and (b) FRONSAC encoding.(a) With Cartesian encoding, the PSF of each voxel follows a different pattern across coils, which allows for separate identification of these voxels in a set of aliased images. (b) With FRONSAC encoding, each voxel also has a unique ghosting pattern (constant across coils), which aids in the separation of voxels. Even within a single coil, the PSFs are linearly independent, which aids in the separation of aliased voxels.

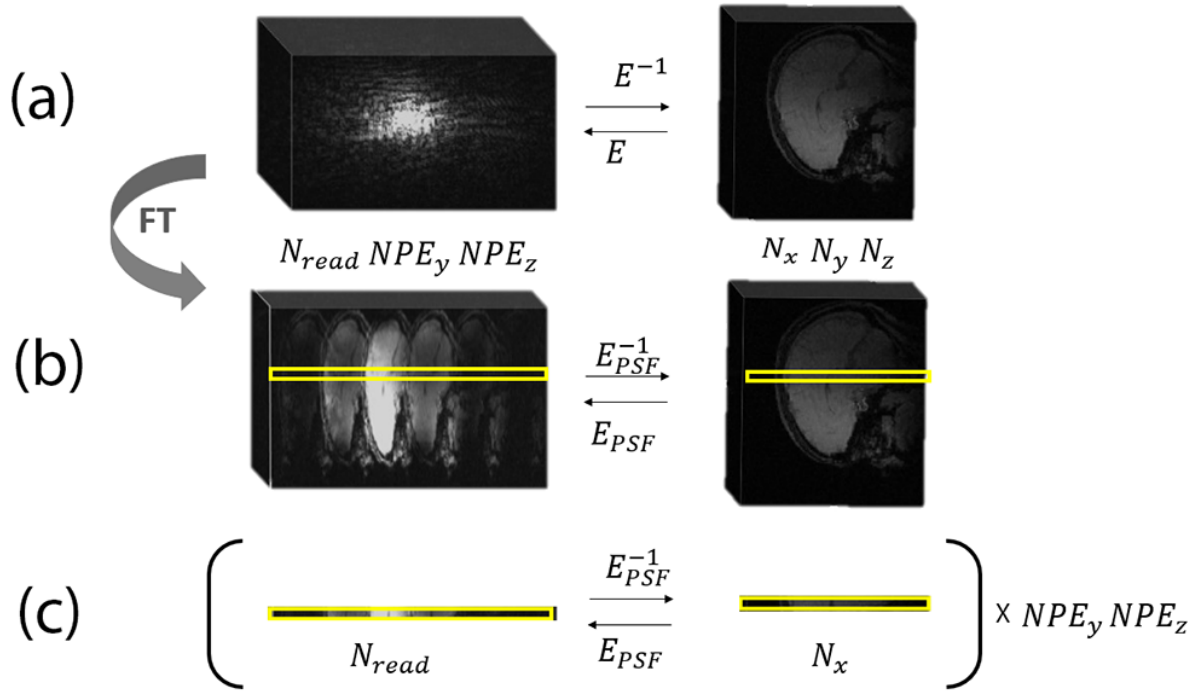

Figure 2: (a) In previous FRONSAC reconstructions, the entire image volume is related to the entire dataset by a very large encoding matrix (generally $N_{read}NPE_yNPE_zN_{coil} \times N_xN_yN_z$ though only one coil is shown in illustration). For most problems, inversion of this matrix, $E$, is not computationally feasible, and even iterative approaches become extremely slow for 3D volumes and high channel counts. (b) After Fourier transform of the data, a far more local relationship is evident, which is analytically derived in the Theory section. This allows one row of data (across coils) to be related to one row (or set of aliased rows) in the reconstruction volume via the matrix $E_{PSF}$. (c) While this smaller matrix inversion must be solved for every row of acquired data, it is equivalent to a diagonalization of the encoding matrix, leading to significantly better efficiency.

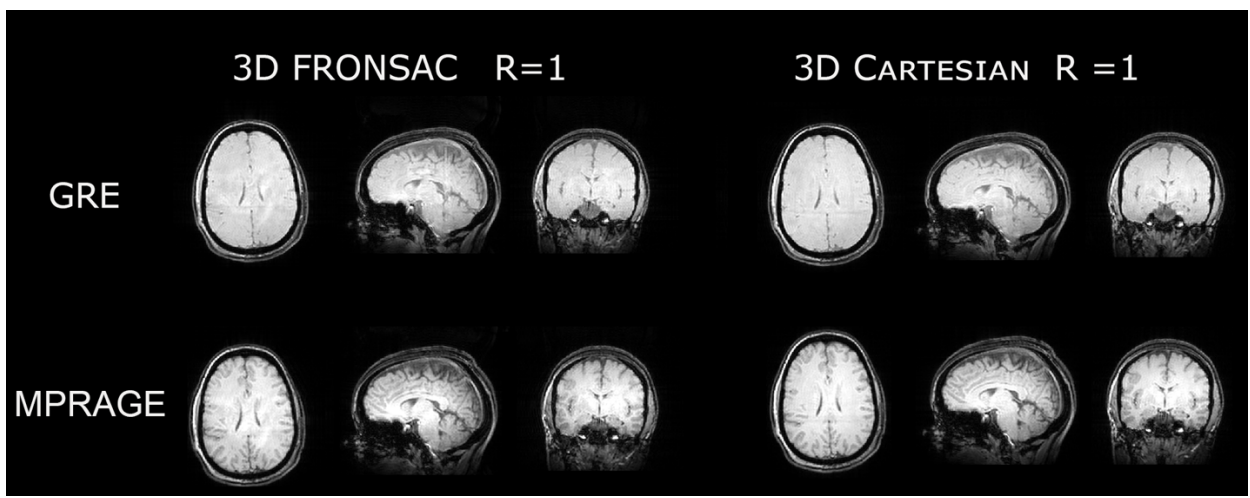

Figure 3: First demonstrations of 3D FRONSAC imaging show very consistent contrast with traditionally acquired GRE and MPRAGE images.

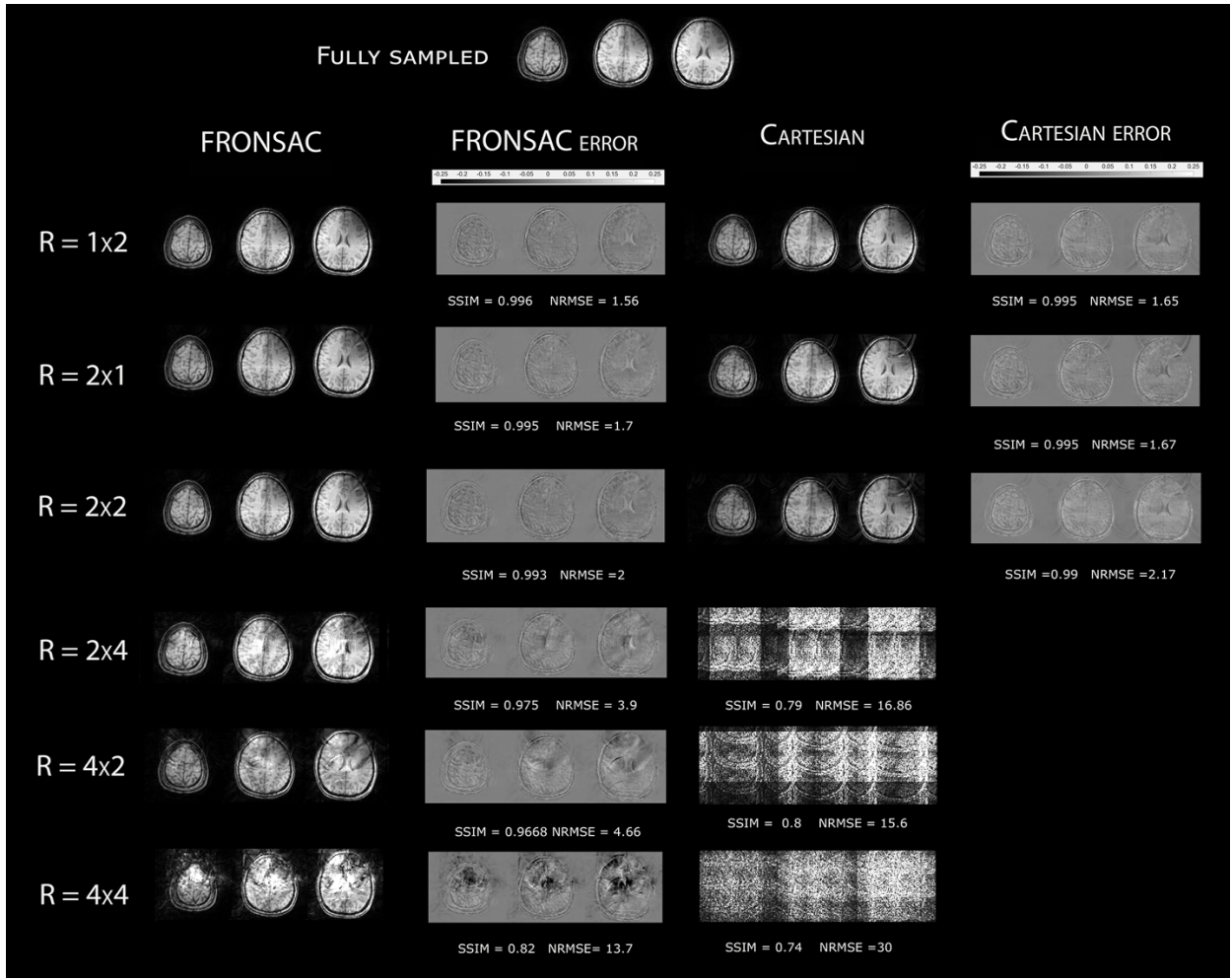

Figure 4: Reconstructions of 32 channel data at various undersampling factors show that struc tural similarity is consistently higher and NRMSE is consistently lower in images enhanced with FRONSAC encoding. In particular, it is notable that the unregularized 2x4 and 4x2 FRONSAC images show relatively modest undersampling artifacts that could possibly be reduced with more sophisticated reconstruction strategies. In contrast, the traditionally encoded Cartesian images fall apart at these high undersampling factors.

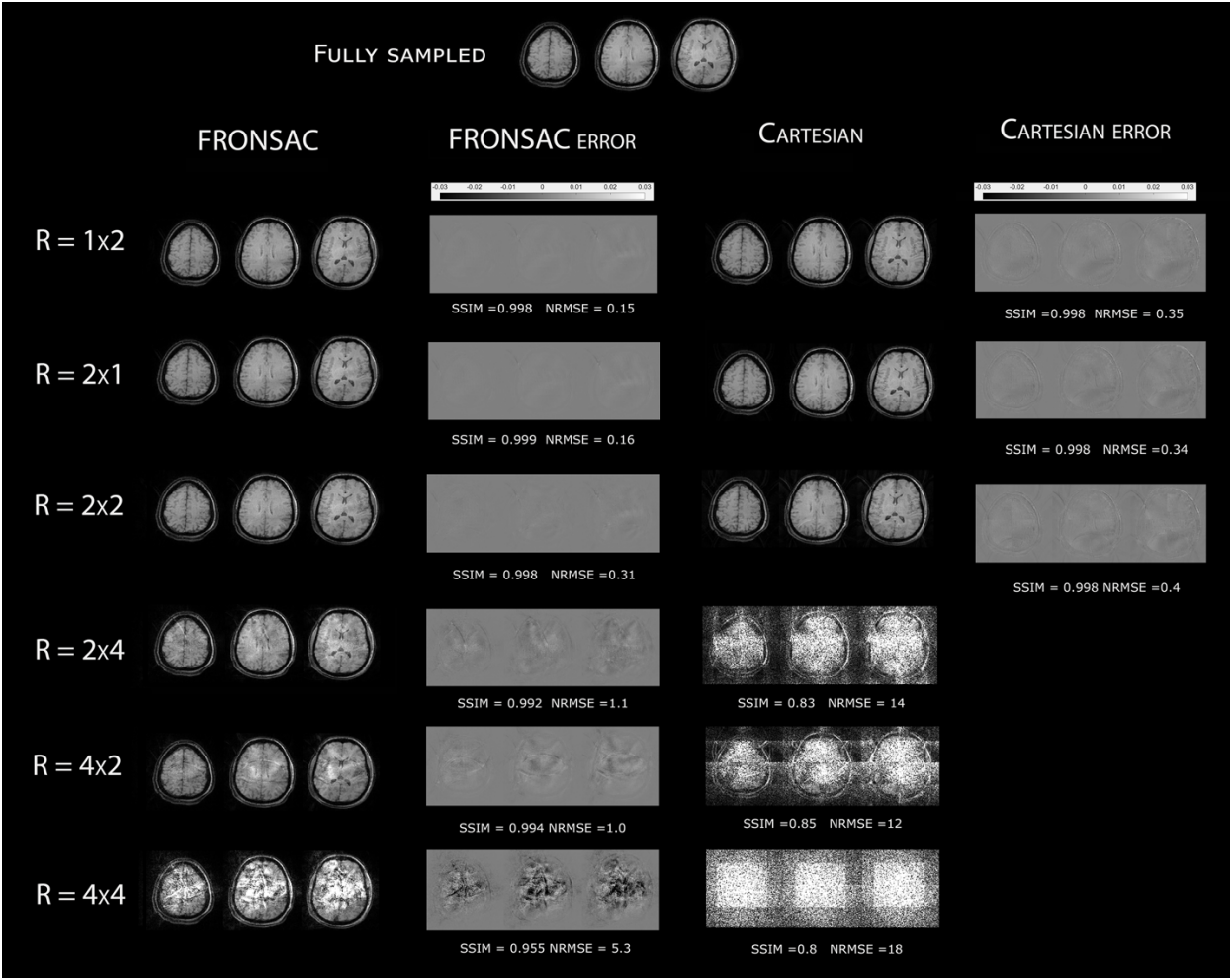

Figure 5: Undersampled reconstructions from a simple 8 channel azimuthally arranged coil show surprisingly good image quality even at relatively high undersampilng factors. As previously described, FRONSAC induces a ghosting pattern for aliased voxels that allows some separation even in the absence of coil encoding. Results are similar to the higher channel reconstructions shown in Figure 4.